\documentclass[prd,onecolumn,nofootinbib]{revtex4}
\usepackage{bm,amsmath,amssymb,graphicx}

\begin{document}

\noindent
International Journal of Modern Physics A {\bf 40} (32), 2544007 (2025)\\

\title{Gravitational collapse with torsion and universe in a black hole\footnote{Talk given at the 14th International Conference in High-Energy Physics  - HEPMAD24 (21-26th October 2024, Antananarivo,\\Madagascar).}}
\author{Nikodem Pop{\l}awski}
\altaffiliation{NPoplawski@newhaven.edu}

\affiliation{Department of Mathematics and Physics, University of New Haven, West Haven, CT, USA}

\begin{abstract}
We consider gravitational collapse of a fluid sphere with torsion generated by spin, which forms a black hole.
We use the Tolman metric and the Einstein--Cartan field equations with a relativistic spin fluid as a source.
We show that gravitational repulsion of torsion prevents a singularity, replacing it with a nonsingular bounce.
Quantum particle creation during contraction prevents shear from overcoming torsion.
Particle creation during expansion can generate a finite period of inflation and produce large amounts of matter.
The resulting closed universe on the other side of the event horizon may have several bounces.
Such a universe is oscillatory, with each cycle larger than the preceding cycle, until it reaches a size at which dark energy dominates and expands indefinitely.
Our Universe might have therefore originated from a black hole existing in another universe.
\end{abstract}
\maketitle

{\bf 1. Einstein--Cartan gravity and spin fluid}\\ \\
In general relativity (GR), the affine connection $\Gamma^{k}_{ij}$ is symmetric \cite{Schr}.
The Einstein--Cartan--Sciama--Kibble (EC) theory gravity removes this constraint by regarding the antisymmetric part of the connection, the torsion tensor $S^k_{\phantom{k}ij}=(1/2)(\Gamma^{k}_{ij}-\Gamma^{k}_{ji})$, as a field.
The total Lagrangian density is $(-1/2\kappa)R\sqrt{-g}+\mathcal{L}_m$, as in GR, where $R$ is the Ricci scalar constructed from the connection $\Gamma^{k}_{ij}$, $\mathcal{L}_m$ is the Lagrangian density of matter, and $g$ is the determinant of the metric tensor $g_{ik}$.

Varying the Lagrangian with respect to the contortion tensor $C^k_{\phantom{k}ij}=S_{ij}^{\phantom{ij}k}+S_{ji}^{\phantom{ji}k}+S^k_{\phantom{k}ij}$ gives the Cartan equations \cite{EC}:
\[
    S_{jik}-S_i g_{jk}+S_k g_{ji}=-\frac{1}{2}\kappa s_{ikj},
\]
where $S_i=S^k_{\phantom{k}ik}$ and $s^{ijk}=2(\delta\mathcal{L}_m/\delta C_{ijk})/\sqrt{-g}$ is the spin tensor of matter.
Varying the Lagrangian with respect to the metric tensor $g_{ik}$ gives the Einstein equations with the Ricci tensor constructed from $\Gamma^{k}_{ij}$.
They can be put into a GR form with the Einstein tensor $G^{ik}$ and the energy--momentum tensor of matter $T^{ik}$, combined with contributions from the spin tensor:
\begin{eqnarray}
    & & G^{ik}=\kappa T^{ik}+\frac{1}{2}\kappa^2\biggl(s^{ij}_{\phantom{ij}j}s^{kl}_{\phantom{kl}l}-s^{ij}_{\phantom{ij}l}s^{kl}_{\phantom{kl}j}-s^{ijl}s^k_{\phantom{k}jl}+\frac{1}{2}s^{jli}s_{jl}^{\phantom{jl}k} \nonumber \\
    & & +\frac{1}{4}g^{ik}(2s^{jlm}s_{jml}-2s_{jl}^{\phantom{jl}l}s^{jm}_{\phantom{jm}m}+s^{jlm}s_{jlm})\biggr).
    \nonumber
\end{eqnarray}

Dirac spinors, representing fermions, couple to torsion through the covariant derivative in the Lagrangian, so their intrinsic angular momentum (spin) is the source of torsion.
At macroscopic scales, they can be averaged and described as a spin fluid:
\[
    s_{ij}^{\phantom{ij}k}=s_{ij}u^k,\quad s_{ij}u^j=0,\quad s^2=\frac{1}{2}s_{ij}s^{ij}\propto n_f^2>0,
\]
where $s^2$ is the averaged square of the spin density and $n_f$ is the number density of fermions.
The terms in the combined energy--momentum tensor, which are quadratic in the spin tensor, do not vanish after averaging:
\[
    G^{ij}=\kappa\Bigl(\epsilon-\frac{1}{4}\kappa s^2\Bigr)u^i u^j-\kappa\Bigl(p-\frac{1}{4}\kappa s^2\Bigr)(g^{ij}-u^i u^j).
\]
The Einstein--Cartan equations for a spin fluid are therefore equivalent to the Einstein equations in GR for an ideal fluid with
\begin{equation}
    \tilde{\epsilon}=\epsilon-\alpha n_f^2,\quad\tilde{p}=p-\alpha n_f^2,
    \label{fluid}
\end{equation}
where $\epsilon$ and $p$ are the thermodynamic energy density and pressure, and $\alpha=\kappa(\hbar c)^2/32$ with $\kappa=8\pi G/c^4$ \cite{iso,ApJ,Gabe}.\\

{\bf 2. Gravitational collapse of a fluid sphere}\\ \\
A centrally symmetric gravitational field is given by the Tolman metric \cite{LL2}:
\begin{equation}
    ds^2=e^{\nu(\tau,R)}c^2 d\tau^2-e^{\lambda(\tau,R)}dR^2-e^{\mu(\tau,R)}(d\theta^2+\mbox{sin}^2\theta\,d\phi^2),
    \label{grav1}
\end{equation}
where $\nu$, $\lambda$, and $\mu$ are functions of a time coordinate $\tau$ and a radial coordinate $R$.
Coordinate transformations $\tau\rightarrow\tilde{\tau}(\tau)$ and $R\rightarrow\tilde{R}(R)$ do not change the form of this metric.
The components of the Einstein tensor corresponding to the metric (\ref{grav1}) that do not vanish identically are \cite{LL2}
\begin{eqnarray}
    & & G_0^0=-e^{-\lambda}\Bigl(\mu''+\frac{3}{4}\mu'^2-\frac{1}{2}\mu'\lambda'\Bigr)+\frac{1}{2}e^{-\nu}\Bigl(\dot{\lambda}\dot{\mu}+\frac{1}{2}\dot{\mu}^2\Bigr)+e^{-\mu}, \nonumber \\
    & & G_1^1=-\frac{1}{2}e^{-\lambda}\Bigl(\frac{1}{2}\mu'^2+\mu'\nu'\Bigr)+e^{-\nu}\Bigl(\ddot{\mu}-\frac{1}{2}\dot{\mu}\dot{\nu}+\frac{3}{4}\dot{\mu}^2\Bigr)+e^{-\mu}, \nonumber \\
    & & G_2^2=G_3^3=-\frac{1}{4}e^{-\lambda}(2\nu''+\nu'^2+2\mu''+\mu'^2-\mu'\lambda'-\nu'\lambda'+\mu'\nu') \nonumber \\
    & & -\frac{1}{4}e^{-\nu}(\dot{\lambda}\dot{\nu}+\dot{\mu}\dot{\nu}-\dot{\lambda}\dot{\mu}-2\ddot{\lambda}-\dot{\lambda}^2-2\ddot{\mu}-\dot{\mu}^2), \nonumber \\
    & & G_0^1=\frac{1}{2}e^{-\lambda}(2\dot{\mu}'+\dot{\mu}\mu'-\dot{\lambda}\mu'-\dot{\mu}\nu'),
    \label{grav2}
\end{eqnarray}
where a dot denotes differentiation with respect to $c\tau$ and a prime denotes differentiation with respect to $R$.

In the comoving frame of reference, the spatial components of the four-velocity $u^i$ vanish.
The nonzero components of the energy--momentum tensor for a spin fluid, $T_{ik}=(\tilde{\epsilon}+\tilde{p})u_i u_k-\tilde{p}g_{ik}$, are: $T^0_0=\tilde{\epsilon}$, $T^1_1=T^2_2=T^3_3=-\tilde{p}$.
The Einstein field equations $G^i_k=\kappa T^i_k$ in this frame of reference are: $G_0^0=\kappa\tilde{\epsilon},\quad G_1^1=G_2^2=G_3^3=-\kappa\tilde{p},\quad G_0^1=0$.
The covariant conservation of the energy--momentum tensor gives \cite{LL2}
\begin{equation}
    \dot{\lambda}+2\dot{\mu}=-\frac{2\dot{\tilde{\epsilon}}}{\tilde{\epsilon}+\tilde{p}},\,\,\,\nu'=-\frac{2\tilde{p}'}{\tilde{\epsilon}+\tilde{p}},
    \label{grav4}
\end{equation}
where the constants of integration depend on the allowed transformations $\tau\rightarrow\tilde{\tau}(\tau)$ and $R\rightarrow\tilde{R}(R)$.

If the pressure is homogeneous (no pressure gradients), then $\tilde{p}'=0$ and $p=p(\tau)$.
In this case, the second equation in (\ref{grav4}) gives $\nu'=0$.
Therefore, $\nu=\nu(\tau)$ and a transformation $\tau\rightarrow\tilde{\tau}(\tau)$ can bring $\nu$ to zero and $g_{00}=e^\nu$ to 1.
The system of coordinates becomes synchronous.
Defining $r(\tau,R)=e^{\mu/2}$ turns the metric (\ref{grav1}) into
\begin{equation}
    ds^2=c^2 d\tau^2-e^{\lambda(\tau,R)}dR^2-r^2(\tau,R)(d\theta^2+\mbox{sin}^2\theta\,d\phi^2).
    \label{grav5}
\end{equation}
The Einstein equations (\ref{grav2}) reduce to \cite{LL2}
\begin{eqnarray}
    & & \kappa\tilde{\epsilon}=-\frac{e^{-\lambda}}{r^2}(2rr''+r'^2-rr'\lambda')+\frac{1}{r^2}(r\dot{r}\dot{\lambda}+\dot{r}^2+1), \nonumber \\
    & & -\kappa\tilde{p}=\frac{1}{r^2}(-e^{-\lambda}r'^2+2r\ddot{r}+\dot{r}^2+1), \nonumber \\
    & & -2\kappa\tilde{p}=-\frac{e^{-\lambda}}{r}(2r''-r'\lambda')+\frac{\dot{r}\dot{\lambda}}{r}+\ddot{\lambda}+\frac{1}{2}\dot{\lambda}^2+\frac{2\ddot{r}}{r}, \nonumber \\
    & & 2\dot{r}'-\dot{\lambda}r'=0.
    \label{grav6}
\end{eqnarray}

Integrating the last equation in (\ref{grav6}) gives
\begin{equation}
    e^\lambda=\frac{r'^2}{1+f(R)},
    \label{grav7}
\end{equation}
where $f$ is a function of $R$ satisfying a condition $1+f>0$ \cite{LL2}.
Putting the relation (\ref{grav7}) into the second equation in (\ref{grav6}) gives $2r\ddot{r}+\dot{r}^2-f=-\kappa\tilde{p}r^2$, which is integrated to \cite{collapse}
\begin{equation}
    \dot{r}^2=f(R)+\frac{F(R)}{r}-\frac{\kappa}{r}\int\tilde{p}r^2 dr,
    \label{grav8}
\end{equation}
where $F$ is a positive function of $R$.
The third equation in (\ref{grav6}) does not give a new relation.
Putting the relation (\ref{grav7}) into the first equation in (\ref{grav6}) and using equation (\ref{grav8}) gives $\kappa(\tilde{\epsilon}+\tilde{p})=F'(R)/(r^2 r')$ \cite{collapse}.
Combining it with equation (\ref{grav8}) gives
\begin{equation}
    \dot{r}^2=f(R)+\frac{\kappa}{r}\int_0^R\tilde{\epsilon}r^2 r'dR.
    \label{grav10}
\end{equation}

Every particle in a collapsing fluid sphere is represented by a radial coordinate $R$ that ranges from 0 (at the center of the sphere) to $R_0$ (at the surface of the sphere).
The relation (\ref{grav10}) is the equation of radial motion for a particle with a given value of $R$.
If the mass of the sphere is $M$, then the Schwarzschild radius $r_g=2GM/c^2$ of the black hole that forms from the sphere is equal to $r_g=\kappa\int_0^{R_0}\tilde{\epsilon}r^2 r'dR$ \cite{LL2}.
Equation (\ref{grav10}) for $R=R_0$ gives the equation of motion for a particle at the surface of the sphere:
\begin{equation}
    \dot{r}^2(\tau,R_0)=f(R_0)+\frac{r_g}{r(\tau,R_0)}.
    \label{grav12}
\end{equation}
If $r_0=r(0,R_0)$ is the initial radius of the sphere and the sphere is initially at rest, then $\dot{r}(0,R_0)=0$.
Consequently, the relation (\ref{grav12}) determines the value of $R_0$:
\begin{equation}
    f(R_0)=-\frac{r_g}{r_0}.
    \label{grav13}
\end{equation}

Putting the relations $r=e^{\mu/2}$ and (\ref{grav7}) into the first equation in (\ref{grav4}) gives $(\tilde{\epsilon}r^2 r')^\cdot+\tilde{p}(r^2 r')^\cdot=0$, which is the first law of thermodynamics for the effective $\tilde{\epsilon}$ and $\tilde{p}$ (\ref{fluid}) \cite{ApJ}.\\

{\bf 3. Collapse of a spin fluid sphere}\\ \\
If we consider a spin fluid, which is composed by an ultrarelativistic matter in kinetic equilibrium, then $\epsilon=h_\star T^4$, $p=\epsilon/3$, and $n_f=h_{nf}T^3$, where $T$ is the temperature of the fluid, $h_\star=(\pi^2/30)(g_b+(7/8)g_f)k_B^4/(\hbar c)^3$, and $h_{nf}=(\zeta(3)/\pi^2)(3/4)g_f k_B^3/(\hbar c)^3$ \cite{ApJ,Gabe}.
For standard-model particles, $g_b=29$ and $g_f=90$.
Therefore, the effective energy density and pressure (\ref{fluid}) depend on the temperature according to
\[
    \tilde{\epsilon}=h_\ast T^4-\alpha h_{nf}^2 T^6,\quad \tilde{p}=\frac{1}{3}h_\ast T^4-\alpha h_{nf}^2 T^6.
\]

If the pressure has no gradient, then the temperature depends only on $\tau$, and so does the energy density.
This case describes a homogeneous sphere.
Putting them into the first law of thermodynamics gives
\begin{equation}
    r^2 r'T^3=g(R),
    \label{spin2}
\end{equation}
where $g$ is a function of $R$.
Putting this relation into equation (\ref{grav10}) gives \cite{collapse}
\begin{equation}
    \dot{r}^2=f(R)+\frac{\kappa}{r}(h_\star T^4-\alpha h_{nf}^2 T^6)\int_0^R r^2 r'dR.
    \label{spin3}
\end{equation}
Equations (\ref{spin2}) and (\ref{spin3}) give the function $r(\tau,R)$, which with the relation (\ref{grav7}) gives the function $\lambda(\tau,R)$.
The integral of the equation of motion (\ref{spin3}) also contains the initial value $\tau_0(R)$.
A general solution of the Einstein equations (\ref{grav6}) for the metric (\ref{grav5}) therefore depends on three arbitrary functions: $f(R)$, $g(R)$, and $\tau_0(R)$.

We seek a solution of equations (\ref{spin2}) and (\ref{spin3}) as
\begin{equation}
    f(R)=-\sin^2 R,\quad r(\tau,R)=a(\tau)\sin R,
    \label{spin4}
\end{equation}
where $a(\tau)$ is a nonnegative function of $\tau$.
For these functions, the relation (\ref{spin2}) gives $a^3 T^3\sin^2 R\cos R=g(R)$, in which separation of the variables $\tau$ and $R$ leads to
\[
    g(R)=\mbox{const}\cdot \sin^2 R\cos R,\quad a^3 T^3=\mbox{const}.
\]
Consequently, we obtain
\begin{equation}
    aT=a_0 T_0,\quad \frac{\dot{T}}{T}+\frac{H}{c}=0,
    \label{spin7}
\end{equation}
where $a_0=a(0)$, $T_0=T(0)$, and $H=c\dot{a}/a$.
Putting the relations (\ref{spin4}) into the equation of motion (\ref{spin3}) gives \cite{collapse}
\begin{equation}
    \dot{a}^2+1=\frac{1}{3}\kappa(h_\star T^4-\alpha h_{nf}^2 T^6)a^2,
    \label{spin8}
\end{equation}
which has the form of the Friedmann equation for the scale factor $a$ as a function of the cosmic time $\tau$ in a closed, homogeneous, and isotropic universe.
The quantity $H$ is the Hubble parameter of this universe.
Using the first relation in (\ref{spin7}) to replace $T$ with $a$ in the Friedmann equation (\ref{spin8}) yields
\begin{equation}
    \dot{a}^2=-1+\frac{1}{3}\kappa\Bigl(\frac{h_\star T^4_0 a^4_0}{a^2}-\frac{\alpha h_{nf}^2 T_0^6 a^6_0}{a^4}\Bigr).
    \label{spin9}
\end{equation}

The values of $a_0$ and $R_0$ are determined by the relations (\ref{grav13}) and (\ref{spin4}) \cite{LL2,collapse}:
\begin{equation}
    \sin R_0=\Bigl(\frac{r_g}{r_0}\Bigr)^{1/2},\quad a_0=\Bigl(\frac{r_0^3}{r_g}\Bigr)^{1/2}.
    \label{size}
\end{equation}
Putting the initial values $a_0$ and $\dot{a}(0)=0$ into equation (\ref{spin8}), in which the second term on the right-hand side is negligible, gives $Mc^2=(4\pi/3)r^3_0 h_\star T^4_0$.
This relation indicates the equivalence of mass and energy of a fluid sphere with radius $r_0$ and determines $T_0$.
An event horizon for the sphere forms when $r(\tau,R_0)=r_g$, which is equivalent to $a=(r_g r_0)^{1/2}$.
Equation (\ref{spin9}) has two turning points, $\dot{a}=0$, if $r^3_0/r_g>(3\pi G\hbar^4 h_{nf}^4)/(8h_\star^3)\sim l_P^2$ ($l_P=$ Planck length), valid for systems forming black holes \cite{Gabe}.

Putting the relations (\ref{spin4}) into the function (\ref{grav7}) gives $e^{\lambda(\tau,R)}=a^2$.
Consequently, the square of an infinitesimal interval in the interior of a collapsing spin fluid (\ref{grav5}) is given by $ds^2=c^2 d\tau^2-a^2(\tau)dR^2-a^2(\tau)\sin^2 R(d\theta^2+\mbox{sin}^2\theta\,d\phi^2)$ \cite{LL2}.
The initial value of the scale factor $a$ is equal to $a_0$.
This metric has a form of the closed Friedmann--Lema\^{i}tre--Robertson--Walker metric and describes a part of a closed universe with $0\le R \le R_0$.\\

{\bf 4. Nonsingular bounce and particle production}\\ \\
Equation (\ref{spin9}) can be solved analytically in terms of an elliptic integral of the second kind \cite{Gabe}, giving the function $a(\tau)$ and then $r(\tau,R)=a(\tau)\sin R$.
The value of $a$ never reaches zero because as $a$ decreases, the right-hand side of equation (\ref{spin9}) becomes negative, contradicting the left-hand side.
All particles with $R>0$ fall within the event horizon but never reach $r=0$.
A singularity is therefore avoided, and replaced with a nonsingular bounce \cite{collapse}.
Positive values of $a$ give finite values of $T$, $\epsilon$, $p$, and $n_f$.

After a bounce, the matter expands on the other side of the event horizon as a new, closed universe (with constant positive curvature).
The quantity $a(\tau)$ is its scale factor.
This universe is oscillatory: $a$ oscillates between the two turning points.
The value of $R_0$ does not change.
A turning point at which $\ddot{a}>0$ is a bounce, and a turning point at which $\ddot{a}<0$ is a crunch.
The universe has therefore an infinite number of cycles (a cycle lasts from a bounce to a crunch, and back).

The Raychaudhuri equation for a congruence of geodesics without four-acceleration and rotation is $d\theta/ds=-\theta^2/3-2\sigma^2-R_{ik}u^i u^k$, 
where $\theta$ is the expansion scalar and $\sigma^2$ is the shear scalar.
For a spin fluid, the last term in this equation is $-\kappa(\tilde{\epsilon}+3\tilde{p})/2$.
The necessary condition for avoiding a singularity in a black hole is thus $-\kappa(\tilde{\epsilon}+3\tilde{p})/2>2\sigma^2$.
For a relativistic spin fluid, $p=\epsilon/3$, this condition gives $2\kappa\alpha n_f^2>2\sigma^2+\kappa\epsilon$.
Without torsion, the left-hand side of this relation would be absent and this inequality could not be satisfied, resulting in a singularity.
Torsion therefore provides a necessary condition for preventing a singularity \cite{Hehl}.
Without shear, this condition is also sufficient \cite{iso}.

The presence of shear opposes the effects of torsion.
The shear scalar $\sigma^2$ grows with decreasing $a$ like $\sim a^{-6}$, like $n_f^2$.
Therefore, if the initial shear term dominates over the initial torsion term, then it will dominate at later times during contraction, and a singularity would form.
To avoid it, $n_f^2$ must grow faster than $\sim a^{-6}$.
Consequently, fermions must be produced in a black hole during contraction.

A quantum production rate of particles in a varying gravitational field is given by
\begin{equation}
    \frac{1}{c\sqrt{-g}}\frac{d(\sqrt{-g}n_f)}{d\tau}=\frac{\beta H^4}{c^4},\quad g=-a^6\sin^4R\sin^2\theta,
    \label{part1}
\end{equation}
where $\beta$ is a nondimensional production rate.
The second equation in (\ref{spin7}) is generalized into \cite{ApJ}
\begin{equation}
    \frac{\dot{T}}{T}=\frac{H}{c}\Bigl(\frac{\beta H^3}{3c^3 h_{nf}T^3}-1\Bigr).
    \label{part2}
\end{equation}
Particle production changes the power law $n_f(a)$:
\[
    n_f\sim a^{-(3+\delta)},
\]
where $\delta$ varies with $\tau$.
Putting this relation into the production rate (\ref{part1}) gives $\delta\sim -a^\delta\dot{a}^3$.
During contraction, $\dot{a}<0$ and thus $\delta>0$.
The term $n_f^2\sim a^{-6-2\delta}$ grows faster than $\sigma^2\sim a^{-6}$ and a singularity is avoided.
Torsion and particle production together reverse the effects of shear, generating a nonsingular bounce \cite{collapse}.\\

{\bf 5. Closed universe in a black hole and inflation}\\ \\
The dynamics of the nonsingular, relativistic universe in a black hole (on the other side of its event horizon) is described by equations (\ref{spin8}) and (\ref{part2}),
with the initial conditions $a(0)=(r_0^3/r_g)^{1/2}$ and $\dot{a}(0)=0$, giving the functions $a(\tau)$ and $T(\tau)$.
The shear appears in equation (\ref{spin8}) as an additional, positive term $\sim a^{-4}$.
When the universe becomes nonrelativistic, the term $h_\star T^4$ changes into a positive term $\sim a^{-1}$.
The cosmological constant appears as a positive term $\sim a^{2}$.

Particle production increases the maximum size of the scale factor $a$ that is reached at a crunch.
A new cycle therefore lasts longer than the previous cycle.
According to the relations (\ref{size}), $R_0$ is given by $\sin^3 R_0=r_g/a(0)$, where $a(0)$ is the initial scale factor, equal to the maximum $a$ in the first cycle.
Because the maximum $a$ in the next cycle is larger, $\sin R_0$ decreases and $R_0\to\pi$ (completely closed universe) \cite{collapse}.

During contraction, $H$ is negative and the temperature $T$ increases.
During expansion, $H$ is positive, so if $\beta$ is too big, then the right-hand side of equation (\ref{part2}) could become positive.
The temperature would then grow with increasing $a$ (eternal inflation).
To avoid it, the production rate has an upper limit: the maximum of the positive function $(\beta H^3)/(3c^3 h_{nf}T^3)$ must be smaller than 1 \cite{ApJ}.

If the function $(\beta H^3)/(3c^3 h_{nf}T^3)$ increases after a bounce to a value slightly smaller than 1, then $T$ becomes approximately constant.
Accordingly, $H$ is also nearly constant and $a$ grows exponentially, generating inflation \cite{ApJ}.
Because $\epsilon$ during inflation is also nearly constant, the universe produces large amounts of matter and entropy.
Such an expansion lasts until this function drops below 1.
Consequently, inflation generated by torsion and particle production lasts a finite period of time, after which torsion weakens and the universe begins the radiation- and then the matter-dominated expansion.

If quantum effects near a bounce do not produce enough matter, then the closed universe reaches a maximum size and then contracts to another bounce, beginning a new cycle.
Because of matter production, a new cycle reaches a larger maximum size than the previous cycle \cite{ApJ}.
When the universe reaches a size at which the cosmological constant dominates, it avoids another contraction and expands to infinity.
The last bounce: the big bounce, is the big bang of the universe created by a black hole.

A parent black hole creating a new, baby universe becomes an Einstein–-Rosen bridge (unidirectional wormhole) to that universe \cite{ER}.
A closed universe is the three-dimensional hypersurface (with radius $a$) of a four-dimensional hypersphere.
Accordingly, our Universe may be closed and born in the interior of a black hole existing in a parent universe \cite{ER}.
Torsion and particle production generate finite inflation, which is consistent with the astronomical data \cite{SD}.
This scenario also occurs if the fermionic matter is described by the Dirac fields instead of the spin fluid \cite{spin}.
It would still be valid for a more realistic gravitational collapse of an inhomogeneous and rotating fluid.

In addition to eliminating gravitational singularities, torsion may also remove divergences in Feynman diagrams in quantum electrodynamics, resulting in finite values of bare (before renormalization) quantities such as the mass and electric charge of the electron \cite{Mike}.
Torsion may be a physical mechanism, ensuring that all physics is finite.

\section*{Acknowledgments}
I am grateful to Francisco Guedes and my Parents, Bo\.{z}enna Pop{\l}awska and Janusz Pop{\l}awski, for their support.

\end{document}